# The recombinase protein is a torque sensitive molecular switch.

*Scott X. Atwell[1], Aurélie Dupont[1,2], Daniel Migliozzi[1], Jean-Louis Viovy[1], Giovanni Cappello[1,2,*]*

[1] Institut Curie, PSL Research University, Centre National de la Recherche Scientifique, Sorbonne Universités, UPMC Univ Paris 06, Unité Mixte de Recherche 168, & institut Pierre Gilles de Gennes, MMBM group 75005 Paris, France

[2] Laboratoire Interdisciplinaire de Physique, Centre National de la Recherche Scientifique, Université Grenoble Alpes, Unité Mixte de Recherche 5588, 38402 St. Martin d'Hères, France

** To whom correspondence should be addressed. Tel: +33 (0)476 514 798; Fax +33(0) 476 635 495; Email: Giovanni.Cappello@univ-grenoble-alpes.fr

## Abstract

**How a nano-searcher finds its nano-target is** a general problem in non-equilibrium statistical physics. It becomes **vital when the searcher is a damaged DNA fragment** trying to find its counterpart on the intact homologous chromosome. If the two copies are paired, that intact homologous sequence serves as a template to reconstitute the damaged DNA sequence, enabling the cell to survive without genetic mutations. To succeed, the search must stop only when the perfect homology is found. The biological process that ensures such a genomic integrity is called **Homologous Recombination** and is promoted by the Recombinase proteins. In this article, we use torque-sensitive magnetic tweezers to measure the free-energy landscape of the human Recombinase hRad51 protein assembled a DNA fragment. Based on our measurements we model the hRad51/DNA complex as an out-of-equilibrium two-state system and provide a thermodynamical description of Homologous Recombination. With this dynamical two-state model, we suggest a mechanism by which the recombinase proteins discriminate between homologous and a non-homologous sequences.

Any chemical reaction happens in two phases: a "searching phase", during which the reactants get in contact and recognize each other, followed by the reaction itself. Whereas the searching phase is reversible, an irreversible and dissipative event, occurring during the second phase, is needed to make the overall reaction irreversible and stabilize the reaction product(s).The nature of this dissipative event is generally clear in simple chemical reactions, and associated with the following of a reaction path with an energy barrier. For biochemical reactions involving several partners with multiple conformational degrees of freedom and spatially extended interactions, however, deciphering the details of the reaction path may be more delicate. This is for instance the case for Homologous Recombination (HR), which happens in living cells to avoid irreversible chromosome damage in case of double-strand breaks. HR uses the redundant genetic information stored in the sister chromatid to accurately reconstruct the damaged DNA. As a major actor of crossing over in meiosis, HR is also fundamental to maintain genetic diversity[1].

Contrarily to chemical reactions, which occur between two pools of ~$10^{23}$ indistinguishable molecules, Homologous Recombination takes place between two unique individual molecules of very large size and complexity: the damaged DNA and its homologous counterpart. The recombination with any other sequence either heterologous or partially homologous must be avoided, as it may lead to genetic instability. Thus, the dissipative step of Homologous Recombination has to occur when, and only when, the homologous sequence is found. Clearly, the free energy associated to Watson-Crick pairing between homologous sequences drives the homology recognition. During this process, the slight energy shift due to the presence of a homology mismatch has to be detected to hinder unwanted recombination. Recent works provide a precise experimental determination of the free energy reduction ΔG during the Recombination-mediated base pairing[2] and evaluate the energy cost ΔΔG associated to a given mismatch[3]. Although the mismatch cost is surprisingly small, it is sufficient to trigger or not the irreversibility of the HR process. Why HR process is so sensitive to such small variations of ΔG is still under debate. Beside these energetic considerations, HR also requires the crossing of a strong activation energy barrier, associated with the base-pairing of the target, intact DNA. Recombinases, and in particular for humans hRad51, play a key role in redefining the reaction path to make it compatible with the energetic quanta associated with individual ATP hydrolysis steps, and imposing the dissipative step making HR irreversible. We investigate here its key role in the sensitivity of HR to mismatches, and thus in the final fidelity of the process.

In order to determine the physical mechanism underlying HR high sensitivity to pairing mismatch, we use new torque-sensitive magnetic tweezers (see schema in Figure 1a) to measure the free energy landscape of the human Recombinase protein hRad51 in interaction with a single double stranded DNA molecule. The active unit of HR is a Recombinase/DNA complex, called Nucleoprotein Filament (*NpF*). The *NpF* is a helical filament, with about six proteins per turn[4]. Each protein binds one ATP molecule and covers three DNA bases along a single strand DNA. Inside the *NpF* the DNA molecule is stretched to 150% of its crystallographic length, and unwound by 43%[5]. Experiments by van Mameren *et al*[6] indicate that in such an extended state, the DNA is likely to be denatured. This stretched filament is also called *active NpF*, as it is able to find the homologous sequence *in vitro* without ATP hydrolysis[7]. After ATP hydrolysis, the *NpF* is converted into the *inactive* (ADP-bound) more compact form. Although the inactive form is not able to achieve homology search, ATP hydrolysis is required to promote the strand exchange and the consequent dismantling of the *NpF*[8].

In summary, there is a clear correlation between the chemical states of the *NpF* (ATP/ADP), its structures (stretched/condensed, unwound/wound), its activity (active/inactive) and, ultimately, its function. As mentioned above, the specificity of HR, and in particular its high sensitivity to base-pair mismatch down to the single base-pair, are key features, and must lie in the detailed thermodynamics of the interactions between recombinases and DNA at the molecular level. It is thus tempting to use the power of single-molecule experiments to validate potential molecular models of this extremely powerful and still elusive mechanism. Unfortunately, constructing a full single-molecule experimental model of the three-strands exchange, while measuring the full set of mechanical parameters (extension, stretching force, torsion and torque), still raises unsolved experimental difficulties. Elegant optical tweezers experiments allowed direct observation of strand pairing in a NpF, but they do not give access to all of the above mechanical parameters, necessary to describe the system's thermodynamics[9,10,11]. In order to nevertheless unravel the physical/thermodynamical origin of the specificity and reliability of strand exchange, we propose here an alternate approach. We use in particular the fact, previously demonstrated, that in the absence of external mechanical constraints, Rad51 adopts the same structure around a single strand and around a double-strand DNA[12]. Thanks to experiments performed on a dsDNA, we can thus measure the intrinsic mechanical features of the protein helix, which would be impossible with ssDNA due to the lack of torsional rigidity of the latter. We combine these newly measured parameters with already known properties of DNA into a two-states model for strand-exchange, and finally compare the predictions of this model with the known properties of Homologous Recombination.

As the homology search phase is reversible, the total energy is conserved during this phase. This implies that the free energy of the three-strand synapse (nucleoprotein filament in contact with both ssDNA and dsDNA) does not depend on the order, by which the ssDNA and the dsDNA molecules interact with the NpF. Within this hypothesis, we evaluate the synapse formation energy by measuring the mechanical work done by the hRad51 protein to stretch and to unwind a dsDNA molecule. This mechanical work is deduced by measuring the torque and the torsion that the hRad51 applies to the dsDNA molecule, as well the extension of the dsDNA molecule under the effect of an external force, during the transition from the condensed to the stretched form. We then model the hRad51 protein as a molecule with two mechanically-coupled conformational states (stretched and condensed), and we determine the free-energy landscape associated to the transition between the two forms. In addition, we detect at which point of the transition the ATP is hydrolyzed to ensure the irreversibility of the recombination process. Taken together, our experimental and modelling work provide a new physical mechanism to explain the high sensitivity of HR to the level of homology between the two DNA sequences. In particular, we show that the mechanical description of the hRad51 protein in terms of a two-state switch explains a non-linear response to small variations ΔΔG around ΔG. Of course, a description of the reaction at the microscopic level is still necessary to precisely describe the reaction kinetics.

## Results

### Mechanically-induced conformational transition of hRad51 nucleoprotein filaments

In a previous work[13], we showed that in the magnetic tweezers the hRad51 protein spontaneously binds to double-stranded DNA molecules (*dsDNA*) with two different conformations. When the tweezers unwinds the dsDNA by 43% as compared to the B-DNA (supercoiling degree $\sigma = -0.43$), the NpF assembles in the stretched conformation (relative extension $L/L_c$ = 1.5, $L_c$ being the crystallographic length of dsDNA). This conformation is also observed when the hRad51 binds to nicked plasmids, which cannot support a torsional stress. Hereafter, we will refer to this state as *S-state*. Conversely, when assembled on dsDNA molecules constrained at the natural topology of B-DNA ($\sigma = 0$) the NpF assembles in a condensed state, called *C-state* in the following. The *C-state* is typically observed on un-nicked plasmids, which do not release the torsional stress.

Figure 1b illustrates the relative DNA elongation in the magnetic tweezers, when the hRad51 proteins are injected into the observation chamber. After injection of hRad51, with the supercoiling degree clamped at $\sigma = 0$ by the tweezers, the molecule is stretched by 5%±1% as compared to the naked DNA (N = 9 molecules, Standard Deviation = 2.5%). When the torsional constraint is progressively released by steps of $\Delta\sigma \simeq -0.05$ (dotted line), we observe a concomitant increase of the NpF length (continuous line). This indicates that the NpF is mechanically allowed to transit from the *C-state* to the *S-state,* by unwinding it from σ = 0 to σ = -0.43.

The extension of the nucleoprotein filament versus its supercoiling degree is summarized in the Figure 2a (filled circles). When ATP hydrolysis is allowed by the presence of $Mg^{2+}$ ions, the $C \leftrightarrow S$ transition is quasi-reversible and always occurs along the straight line $L(\sigma)$. This means that the relative extension and the supercoiling degree are mechanically coupled.

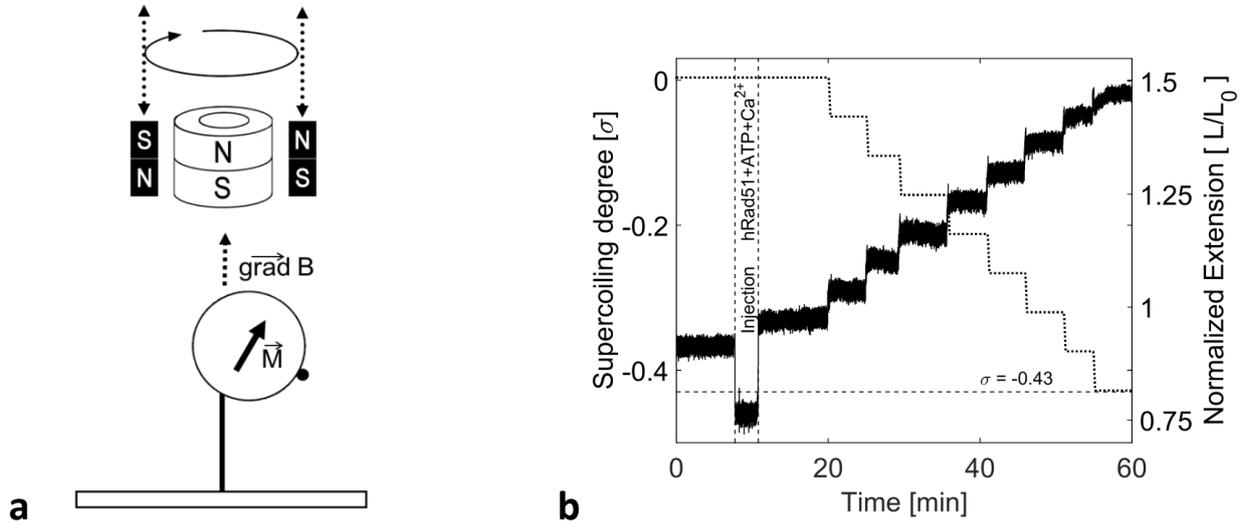

**Figure 1:** (a) Schema of hybrid magnetic tweezers. Small magnets on either side of the main hollow cylindrical magnet are mobile in the vertical direction and able to rotate. The magnetization moment aligns with the quasi vertical magnetic field which also has a small horizontal component. The field gradient remains vertical. (b) Transition from the condensed to the stretched state of the hRad51-dsDNA nucleoprotein filament in ATP and $Mg^{2+}$ through torsion modification. The relative extension of the nucleoprotein filament (continuous line) is shown versus torsion (dotted line). A pulling force higher than 1.4 pN is maintained throughout the whole experiment.

## The condensed nucleoprotein filament is a strained and metastable ADP-bound state

When ATP hydrolysis is impeded by the presence of $Ca^{2+}$ ions[14], the nucleoprotein filament can also assemble in the *C-state*. From this state, it undergoes a transition to the *S-state* when unwound. However the transition is irreversible and, when rewound to $\sigma = 0$ (Figure 2a, empty triangles), the NpF remains partially stretched with a relative extension $L/L_C$ = 1.34±0.03 (N = 5 molecules, standard deviation ±0.05). Eventually, as soon as $Ca^{2+}$ ions are replaced by $Mg^{2+}$ ions the nucleoprotein filament spontaneously shrinks to the *C-state* (Figure 2b). This proves that the $S \rightarrow C$ transition requires ATP hydrolysis.

To clarify the nature of the *C*-state, we directly assemble the NpF in the presence of ADP and without any torsional constraint. When assembled in ADP, the NpF is stable and exhibits a relative extension $L/L_C$ = 1.20±0.05 (see Supplementary Information). Such a relative extension corresponds to $\sigma = -0.15$ and is compatible with the helical pitch of 76Å ($L/L_C$ = 1.16) measured by Yu *et al*[4] in ADP filaments. Thus, we believe that the *C-state* measured here is not the spontaneous ADP-state of the hRad51 protein, but it is forced by the magnetic tweezers that impose the null topology ($\sigma = 0$). *In vivo* the role of magnetic tweezers may be played by the second DNA molecule that has to unwind to embrace the *NpF* conformation. In returns, it induces a torsional stress on the *NpF* and may occasion the *S→C* transition.

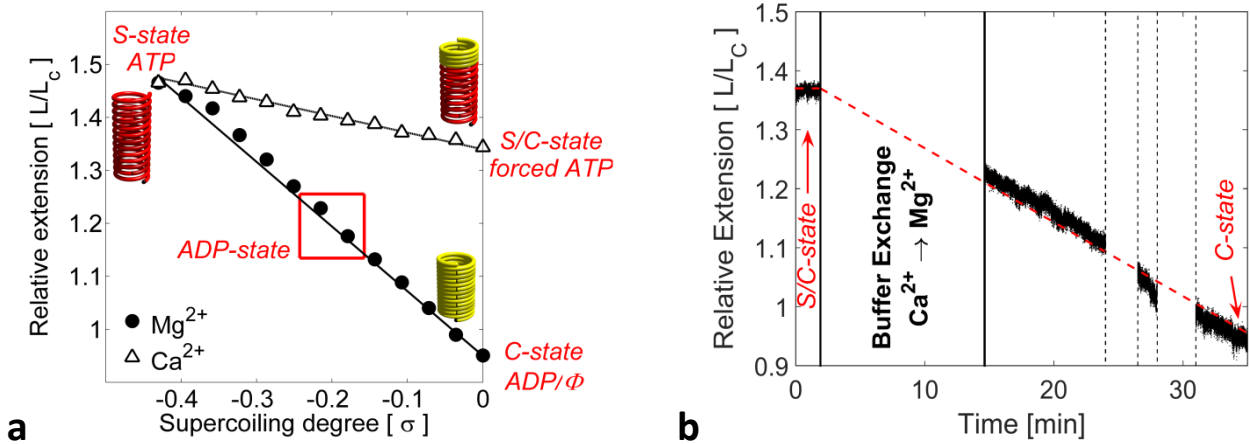

Figure 2: (a) In presence of calcium, the filament remains in the partially extended conformation *S/C* at null torsion ($\sigma=0$), whereas in the presence of magnesium the filament adopts the condensed conformation *C* when wound to $\sigma=0$. (b) Transition from S/C-state to C-state at null-torsion. The gradual shortening of the filament was triggered by buffer exchange $Ca^{2+} \to Mg^{2+}$ and the consequent hydrolysis of ATP. During the buffer exchange (minutes 2-12) the DNA length is not measurable, due to the flow in the chamber. Similarly, the proximity of residual unbound magnetic beads occasionally hinders the tracking (minutes 24-26 and 28-31)

Taken together, these results indicate that the mechanical conformation of the nucleoprotein filament and its chemical state are strongly correlated. In particular, our results show that the stretched state is energetically more favorable in ATP than the condensed one. Conversely, the condensed filament only exists in ADP, as it cannot be reached through a mechanical strain when ATP hydrolysis is impeded.

Noticeably, the *C-state* is metastable. The *NpF* progressively disassembles when forced in the *C-state* by turning the magnetic tweezers to $\sigma = 0$. The disassembly occurs even faster at positive supercoiling degrees, ($\sigma > 0$).

## Torque developed by the Nucleoprotein Filament in the *C→S* transition

During the *C→S* transition the torque applied by the NpF to the dsDNA fragment is elastically transmitted to the bead. The bead rotates clockwise with a typical speed of 1-2 revolutions per minute in the presence of ATP and $Ca^{2+}$. To evaluate the work done by the NpF during the *C→S* transition, we measure the torque in two different manners. First, by progressively lowering the position of the side-magnets, we increase the torque applied to the bead until the rotation is arrested. Figure 3a displays the revolutions of the bead as a function of time. At low external torques ($\Gamma$ < 17 pN·nm) the bead rotation is almost unaffected, whereas it stalls above 24 pN·nm. At intermediate torques (17-22 pN·nm) the rotation is partially arrested after every revolution, when the bead dipole and the magnetic field are aligned, but the NpF still takes advantage of the thermal fluctuations to cross the energy barrier and complete a revolution.

The second method is inspired by Bryan *et al*[15]. The torque developed by hRad51 proteins is deduced from the angular velocity of the bead, $\omega$, during the transition and in the absence of any external magnetic torque (no side-magnets). As explained in S.I., the drag coefficient $\xi$ of the rotating bead is directly inferred from its angular fluctuations. Thus, we deduce the NpF torque $\Gamma(\sigma) = \xi \cdot \omega$ by measuring the angular velocity of the bead at different supercoiling degrees between $\sigma$ = -0.43 and $\sigma$ = 0 (Figure 3, red portions). At one revolution per minute, the whole *C* → *S* transition takes more than 10 hours, which is longer than the NpF dissociation time. Therefore, we use the auxiliary magnets to gradually unwind the filaments (Figure 3b, black segments) and, thus, measure the torque at different supercoiling degrees. We observe that the tendency to unwind held true up to approximately σ=-0.43 corresponding to the supercoiling associated to the *S-state* of the NpF. Depending on the supercoiling degree of the nucleoprotein filament, the measured torque ranges between 5±1 and 25±1 pN·nm (N = 26; standard deviation: 3.5 pN·nm) as shown in Figure 3c. In the presence of $Mg^{2+}$ (ATP hydrolysis is permitted) the maximal measured torque is 21±1 pN·nm (N = 24; standard deviation: 3.5 pN·nm). These values are compatible with those previously measured by Lee *et al*[16].

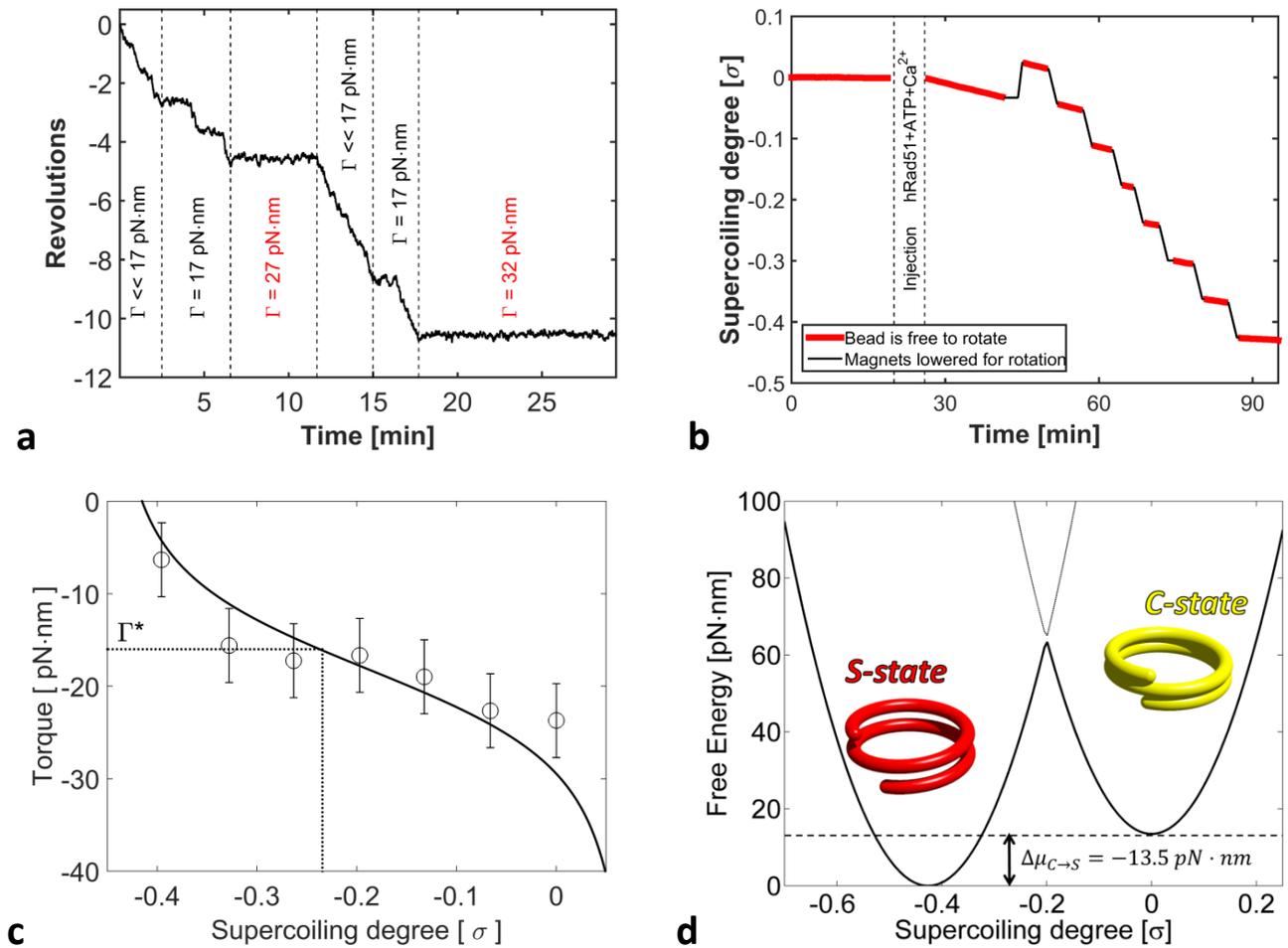

**Figure 3:** (a) Stall torque of the hRad51 NpF. For very small magnetic torques (non-measurable), the filament drives the bead and the rotation is continuous within the experimental time resolution. At 17 pN.nm, pauses appear. The stronger the torque, the longer are the pauses. The pauses occur when the bead dipole is aligned with the external magnetic field. Above ~24 pN the rotation is completely stalled. (b) Rotation of the nucleoprotein filament as a function of time, at different supercoiling degrees. Along the red portions of the curves, the horizontal magnetic field was kept << 17 pN.nm by raising the lateral magnets. To quickly change the supercoiling degree of the filament (black portions of the curve) we force an accelerated bead rotation using the magnetic tweezers. (c) Torque developed by the nucleoprotein filament as a function of its normalized torsion σ. The best fit yields a critical torque $\Gamma^* = 16 \, pN \cdot nm$, for which half of the proteins are in the *S-state*. This corresponds to a difference of chemical potential $\Delta\mu_{C \to S} = -13.5 \pm 1 \, pN \cdot nm$. For a relative extension of 1.05±0.025 (see Figure 1b after hRad51 injection, σ = 0), meaning that 10% of the hRad51 are in the *S-state*, the free-energy is minimal for a torsional stiffness $K_\vartheta = 400 \pm 200 \, pN \cdot nm$. (d) Free Energy associated to the two conformational states of a single hRad51, deduced using the measured torque, force, and stiffness's. For this value of $K_\vartheta$ a barrier of ~45 pN·nm appears between the *C-state* and the *S-state*.

# Two-state mechanical model of hRad51 nucleoprotein filament

Both electron microscopy observations[13] and magnetic tweezers experiments suggest that the hRad51 assembles on the dsDNA either in the condensed or stretched state. Thus, we assume that only those two states exist and propose a quantitative two-state elastic model of the NpF. Using this model, we estimate the difference of internal energy $\Delta\mu$ that drives the spontaneous $C \rightarrow S$ transition in the absence of external stress.

In the following, we consider the NpF as a polymer made of $N$ hRad51 monomers, among which $n_s$ are in the *S-state*. (see Figure SI1). When perturbed by torsion or traction, the hRad51 protein helix either deforms elastically or switch from one state to the other ($S \leftrightarrow C$). For small mechanical strains, we hypothesize that the NpF responds linearly, with torsional and longitudinal stiffness's $K_\vartheta/N$ and $K_\ell/N$ respectively, $K_\vartheta$ and $K_\ell$ being the single protein contribution to the torsional and longitudinal stiffness of the helix.

Within this framework, the free energy is the sum of the elastic energy stored in the strained NpF (extension/torsion), the energy injected by the tweezers, the entropic contribution due to the fact that $n_S/N$ hRad51 are in the *S-state*, and the energy $n_S \cdot \Delta\mu$ released by the $n_S$ proteins in the C$\rightarrow$S transition:

$$G(L,\Theta,n_\mathrm{S}) = \overbrace{\frac{K_\ell}{2N}[L-(N\ell_C+n_\mathrm{S}\Delta\ell)]^2}^{NpF\ Extension\ (elastic)} + \overbrace{\frac{K_\vartheta}{2N}[\Theta-(N\vartheta_C+n_\mathrm{S}\Delta\vartheta)]^2}^{NpF\ Torsion\ (elastic)} - \overbrace{F\cdot L - \Gamma\cdot\Theta}^{Tweezers}$$
$$+ \underbrace{Nk_BT\left[\frac{n_\mathrm{S}}{N}\ln\left(\frac{n_\mathrm{S}}{N}\right) + \left(1-\frac{n_\mathrm{S}}{N}\right)\ln\left(1-\frac{n_\mathrm{S}}{N}\right)\right]}_{Entropy} + \underbrace{n_\mathrm{S}\cdot\Delta\mu}_{Chemical\ Pot.}$$

(1)

Here $L$ and $\Theta$ are respectively the full extension and the torsional angle of the NpF under a force $F$ and a torque $\Gamma$. $\ell_C$ and $\vartheta_C$ are respectively the extension and the torsion of hRad51 in the *C-state*, which correspond to those of B-DNA. Each protein in the *S-state* evokes an additional extension $\Delta\ell$ and a negative torsion $\Delta\vartheta$.

The elastic stiffness $K_\ell$ of a single hRad51 protein bound to dsDNA may be evaluated using the force-extension curve measured by van Mameren *et al.*[17] to $K_\ell = 800 \pm 60\ pN\cdot nm^{-1}$. We are thus left with two unknown parameters: $K_\theta$ and $\Delta\mu_{C\rightarrow S}$ that have to be deduced from our experiments. All the parameters introduced in equation (1) are summarized in Table 1.

### *NpF* torsional stiffness and Energy released in the *C→S* transition.

The partial derivatives of the free energy $G(L, \Theta, n_S)$ with respect to $L, \Theta$ and $n_S$ vanish at the equilibrium state. From equation (1), we derive that (see details in SI):

$$\begin{cases} \frac{n_S}{N} = \frac{1}{1+e^{-(\Gamma \cdot \Delta\vartheta + F \cdot \Delta\ell - \Delta\mu)/k_B T}} \\ \frac{L}{N} = \frac{F}{K_\ell} + \ell_C + \frac{n_S}{N}\Delta\ell \\ \frac{\Theta}{N} = \frac{\Gamma}{K_\vartheta} + \vartheta_C + \frac{n_S}{N}\Delta\vartheta. \end{cases}$$

(2)

In Figure 1b, we show that the dsDNA exhibits a spontaneous elongation of *5±1%* (*L/L°* = 1.05) when it gets in contact with hRad51 proteins, if its topology is kept locked to σ = 0. From the second expression of equation (2), we deduce that this length corresponds to *n_S/N* = 9±2% of hRad51 proteins in the *S-state*. Remarkably, at low force, the value of *n_S/N* only slightly depends on the elastic constant of the protein $K_\ell$, which has been semi-quantitatively deduced from the publication by Van Mameren *et al*[17]. With a stretching force F = 2 pN, a variation of $K_\ell$ by a factor of four leads to a change smaller than 0.8% in the estimation of *n_S/N*.

The ratio *n_S/N* is thermodynamically determined by the balance between the energy *Δμ* released by one hRad51 molecule undergoing the *C→S* transition and the elastic energy stored in the filament of stiffness $K_\vartheta$ during the transition, if the latter does not release the torsional stress (the tweezers impose σ = 0). By combining equations (1) and (3) (see SI) and considering that *n_S/N* = 9±2% at σ = 0, we deduce a torsional stiffness $K_\vartheta$ = 400±150 pN·nm. This stiffness is compatible with that previously measured by Lee *et al*[16] and only slightly depends on the absolute value *Δμ* (an error of ±3 pN·nm in *Δμ* affects $K_\vartheta$ by less than 50 pN·nm).

We precisely determine *Δμ* from the torque *Γ(σ)* developed by the nucleoprotein filament at different supercoiling degrees (Figure 3c). The analytical function *σ(Γ)* is obtained by combining the first and the third expression of equation (2) :

$$\sigma(\Gamma) = \frac{\Theta}{N\vartheta_C} - 1 = \frac{1}{\vartheta_C}\left[\frac{\Gamma}{K_\vartheta} + \frac{\Delta\vartheta}{1+e^{-(\Gamma \cdot \Delta\vartheta + F \cdot \Delta\ell - \Delta\mu)/k_B T}}\right]$$

(3)

Both *Δμ* and $K_\vartheta$ can be estimated by fitting equation (3) to the experimental data. However, $K_\vartheta$ has a little impact on Torque-Supercoiling curve (Figure 3c) and, in practice, an uncertainty of ±200 pN·nm on $K_\vartheta$, gives an error of ±0.6 pN·nm on *Δμ*.

By recursively fitting equation (2b) the Extension-Supercoiling curve (Figure 1b) and equation (3) to the Torque-Supercoiling curve (Figure 3c), we obtain the best agreement for $K_\vartheta$ = 390±110 pN·nm and $\Delta\mu$ = -13.5±1 pN·nm.

## Critical torque of the S➔C transition

The first expression of equation (2) indicates that there is a critical torque $\Gamma^* = -\frac{F \cdot \Delta\ell - \Delta\mu}{\Delta\vartheta}$, at which the hRad51 proteins are half in the *S-state* and half in the *C-state*. In terms of energy, $\Gamma^*$ is the torque at which the mechanical energy introduced by the tweezers $\Delta W^* = \Delta\vartheta \cdot \Gamma^*$ exactly compensates the difference of internal energy $\Delta\mu$. With our values of $K_\vartheta$ and $\Delta\mu$ we find $\Gamma^* \approx 16\ pN \cdot nm \cdot Rad^{-1}$.

For a little perturbation of torque around $\Gamma^*$, the proportion of hRad51 proteins $n_S/N$ in the *S-state* varies very quickly. In this perspective, the NpF is a mechano-sensitive complex, whose maximal sensitivity is at $\Gamma^*$: slightly above $\Gamma^*$ most of the hRad51 proteins undergo the S➔C transition; just below they remain in the *S-state.*

## Energy landscape of the Nucleoprotein Filament

In Homologous Recombination, the nucleoprotein filament is first assembled on a single-stranded DNA molecule. Then, it gets in contact with the homologous double-stranded DNA to initiate the homology search and, eventually, the strand exchange. Even though non-physiological, all the experiments shown above are performed on nucleoprotein filaments directly assembled onto a dsDNA. Indeed, this is the only way to apply torsion to the nucleoprotein filament.

As during homology search the ATP is not hydrolyzed (the reaction remains reversible) the total free-energy is conserved during this phase. As stated before, the free-energy conservation implies that the energy of the three-strand synapse (nucleoprotein filament in contact with both ssDNA and dsDNA) does not depend on the order by which the ssDNA and the dsDNA molecules interact with the NpF. In practice, the energy difference $\Delta\mu^{ss+ds}$ between *C-state* and *S-state* in the three-strand synapse can be written as:

$$\Delta\mu^{ss+ds} = \overbrace{\Delta\mu^{ds} + \Delta W^{ss}}^{dsDNA\ first} \overset{def}{=} \overbrace{\Delta\mu^{ss} + \Delta W^{ds}}^{ssDNA\ first},$$

(4)

where $\Delta\mu^{ss}$ and $\Delta\mu^{ds}$ are the energy difference between *C-state* and *S-state* of a nucleoprotein filament assembled on a ssDNA and on a dsDNA respectively, and $\Delta W^{ds}$ and $\Delta W^{ss}$ the work required to extend and unwind the dsDNA and the ssDNA in the NpF.

When the NpF is assembled on the dsDNA molecule, we measure an energy difference $\Delta\mu^{ds}$ = -13.5 pN·nm (see Figure 3d). Bustamante *et al*[18] estimate $\Delta W^{ss}$ to be around 5 pN·nm, which leads to $\Delta\mu^{ss+ds} = \Delta\mu^{ds} + \Delta W^{ss} \simeq -8.5 \text{ pN} \cdot \text{nm}$. This negative value means that in the three strand synapse the *S-state* remains favorable as compared to the *C-state*, even considering the additional energy required to extend also the third DNA strand (ssDNA). With $\Delta\mu^{ss+ds} = -8.5 \text{ pN} \cdot \text{nm}$, 90% of hRad51 remain in the *S-state*, as illustrated in Figure 4 (b).

From the right hand of equation (4) and with $\Delta W^{ds}$ estimated to about *33 pN·nm* (Léger *et al* [19]), we evaluate the energy needed to compact the nucleoprotein filament assembled on ssDNA molecule $\boldsymbol{\Delta\mu^{ss} \simeq -41.5\ pN \cdot nm}$. With such a large energy difference, the stretched form of the nucleoprotein filament is extremely favored (see Figure 5a, black line), as consensually reported in the literature.

Using the first expression of equation (2), we estimate the ratio of hRad51 protein in the *S-state*, as well as the extension and the topological state of the nucleoprotein filament as a function of the mechanical energy brought by the two DNA molecules (see Figure 4). Whereas the filament is fully stretched in the presence of the ssDNA molecule (point "a", in Figure 4), the critical free energy at which $n_S/N$ *=1/2* is almost reached when the dsDNA molecule is added (b) to form the three strand synapse.

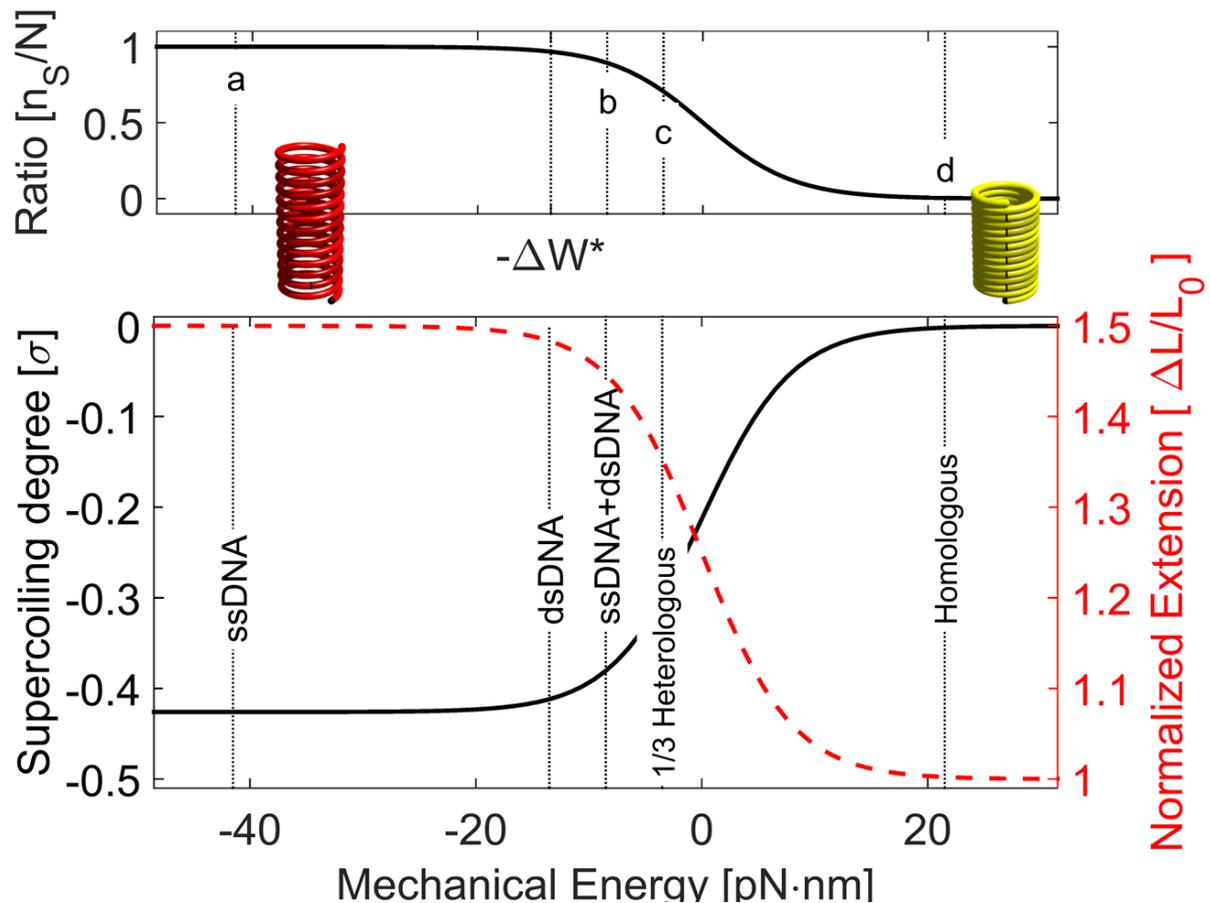

Figure4: Ratio of hRad51 proteins in the S-state, as a function of the energy required to interact with the DNA fragments. Zero Energy it chosen where S-state and C-state are equally probable. In interaction with both ssDNA and dsDNA, the S-state is statistically favored ($n_S/N$ > 90%). Conversely, the S-state becomes penalized ($n_S/N$ < 0.5%) when the full homology is found ("Homologous", "d"). In this case, the filament is almost fully converted in the ADP state. Interestingly, one single mismatch per hRad51 ("1/3 Heterologous", "c") costs enough energy to impede the transition into the condensed state and maintains the Nucleoprotein Filament in a reversible state. Below: supercoiling degree (black) and normalized extension (red) of the Nucleoprotein Filament.

## High sensitivity and irreversibility

The main result of the two-state model is that the NpF responds in a non-linear manner to the stress due to the compression and torsion induced by the ssDNA and dsDNA molecules. At the critical torque (or critical energy) $\Gamma^*$, the response of the nucleoprotein filament is maximal and a little external energy difference has a strong influence on the direction of the $C \leftrightarrow S$ transition. In Homologous Recombination, there is a little energy difference between pairing a perfectly matching sequence and one containing a mismatch. Whereas the pairing of a homologous sequence brings 8 to 12 pN·nm per base-pair[20,21], the denaturation bubble due to a single mismatch gives an energy penalty of about -10 pN·nm[22]. The latter case (single mismatch in a triplet) corresponds to the energy landscape shown by the black curve in Figure 5c, where the two states are equally probable (point "c" in Figure 4) and the supercoiling degree is $\sigma \approx -0.3$. Noticeably, at $\sigma \approx -0.3$ we never observe the depolymerization, as this is below the supercoiling degree associated to the stable ADP-state of the nucleoprotein filament ($\sigma \approx -0.15$). As a consequence, the reaction remains reversible if there is no full homology because the total energy is not sufficient to incorporate the second DNA molecule.

Conversely, the case of full homology corresponds to the energy landscape shown in Figure 5d (black curve). Here the *S→C* transition is thermodynamically favorable and of 98-99.9% of the hRad51 proteins switch in the *C-state* (point "d" in Figure 4). The ATP hydrolysis associated to the *S→C* transition (experiment shown in Figure 2) leads to the consequent of the NpF[23]. If the last part of the reaction happens faster than the reverse transition *C→S* (see Discussion), the reaction becomes irreversible and the final products are (old ssDNA + new dsDNA) are delivered.

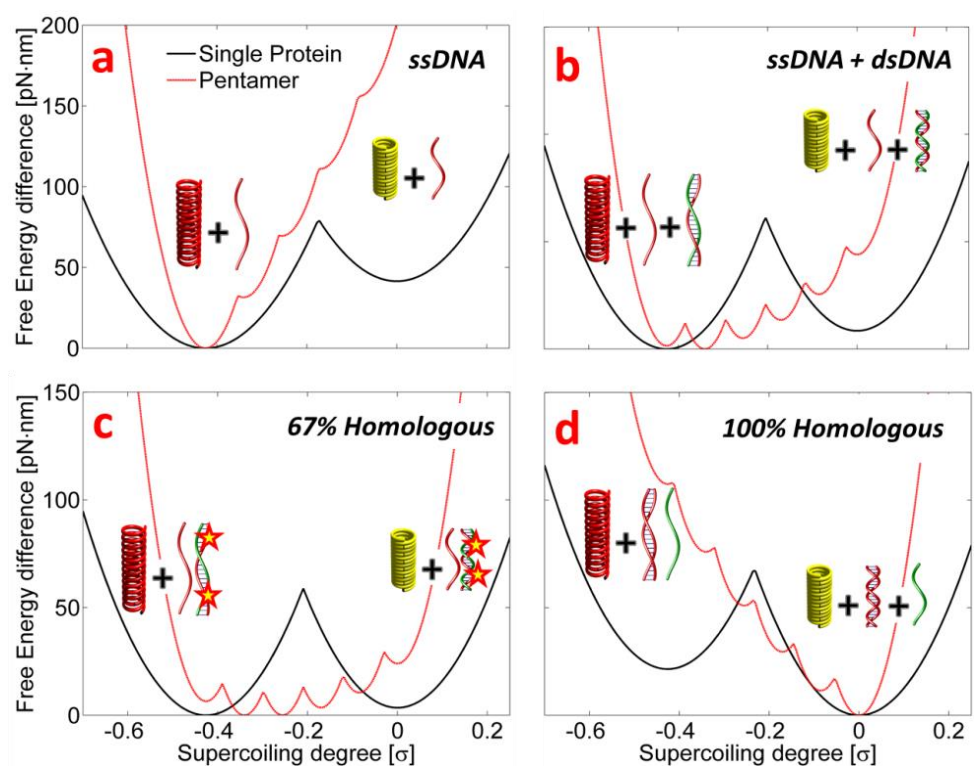

**Figure 5: Free-Energy Landscape of the transition between S-state and C-state in different conditions. The energy is normalized to one hRad51 unit, associated to three base-pairs. (a) Nucleoprotein filament assembled on ssDNA molecules. (b) NpF in interaction with three DNA strands (synapse before strand invasion). (c) Free-Energy landscapes after strand invasion with 67% and (d) with perfect homology. The black curves correspond to the free energy associated to one single unit, while the red curves correspond to the free energy required to compact a pentamer.**

## Activation barrier and homology length selectivity

There is strong evidence for a cooperative behavior of hRad51. In the absence of DNA, hRad51 partially assemble in oligomeric rings. Also in the presence of ssDNA, NpF nucleation requires the simultaneous binding of 4-6 hRad51[24,25] and its depolymerization happens by unbinding of several hRad51 proteins at a time[23]. All these findings suggest that the active unit of HR is not the single hRad51, but a complex of 4-6 proteins.

In terms of mechanics, a complex of several proteins is intrinsically more compliant than a single protein. As a toy model, one needs two times less energy to stretch two springs in series by a length Δx, as compared to a single string of same stiffness. In fact, the strain $Δx/x_0$ of the double spring is twice smaller for a given Δx. This plays an interesting role in our two-state model. As schematized in Figure 6a, a given extension of a *two-state spring* can be reached either in the *S-state* or in the *C-state*, but with different internal mechanical energies. Thermodynamically, the system would evolve toward the state of low energy. However, the mechanical stress accumulated in the spring creates an energy barrier $ΔG_1^*$ between the two states, which hinders (or slows down) the transition. Interestingly, the enhanced compliance of a "dimer" (Figure 6b) has the effect of lowering the elastic barrier to $ΔG_2^* = ΔG_1^*/2$ and facilitates the transition of the single units, one by one. The barrier vanishes for an increasing number of "two-state springs" in series.

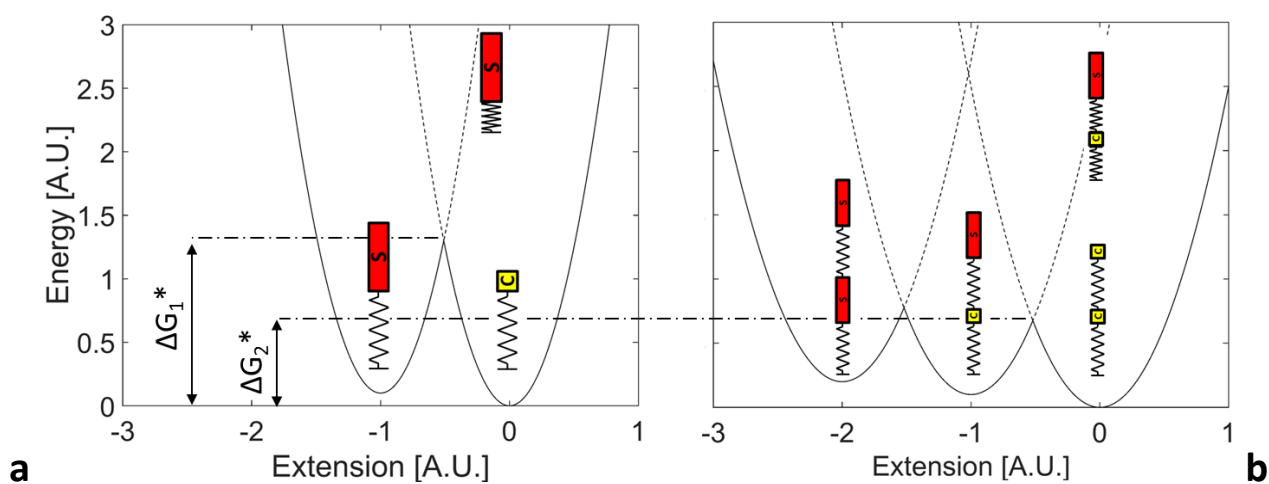

**Figure 6: Potential energy of a two-state spring (A) compared to the energy of 2 two-state springs in series (B). Whereas in both cases *C-state* is energetically more favorable than *S-state*, the height of transition barrier ΔG\* decreases inversely to the number of two-state springs.**

Translated to hRad51 proteins, we use the two-state model to compute the barrier height, both for a single protein and for a pentamer. With a torsional stiffness $K_\vartheta = 400 pN \cdot nm$, $ΔG_1^*$ is in the order of 50 pn·nm (see black line in Figure 5). Such a barrier (12.5 $k_B T$) is high compared to the thermal energy and the dynamics of the transition may be slow in comparison to the lifetime of a three strand synapse, also in case of perfect homology. On the contrary, the Free Energy Landscape of a hRad51-pentamer (red curves in Figure 5) shows five activation barriers, but their height $ΔG_5^*$ almost vanishes in case of perfect homology (Figure 5d). In the latter case, the *S→C* transition occurs instantaneously once the homology is found and the reaction becomes immediately irreversible.

The above model could provide a quantitative explanation of the recent experiments reported by Qi *et al*[2]. The authors point the fundamental role of microhomologies (8-15 consecutive homologous base-pairs) in facilitating the homology recognition and the exchange process. The fact that HR requires 15 consecutive homologous base pairs (3-5 hRad51 adjacent proteins) to succeed, and never happens with a single homologous triplet, is commonly interpreted as an evolutionary way to avoid homology traps. The presence of activation barriers, whose height is inversely proportional to the number of hRad51 units involved in the process, explains why the functional cooperativity of the hRad51 complex penalizes the short homologies and why Homologous Recombination requires 8-15 consecutive homologous base-pairs to succeed.

## Discussion

We measured that the NpF torsional Young's modulus ($K_\vartheta = 390 \pm 110\, pN \cdot nm$) is two orders of magnitude smaller than the typical torsional stiffness of globular proteins[26]. It is notable that the NpF made of RecA protein, the bacterial homolog of hRad51, also exhibits a relatively small torsional stiffness[36]. Interestingly, the torsional stiffness of the NpF is very close to that of bare dsDNA, which is in the range of 300-480 pN·nm[27]. Such a low torsional stiffness of the NpF suggests that its torsional stress is indeed mainly stored in the dsDNA molecule and not in the hRad51 structure. With this hypothesis, hRad51 proteins would only stretch the DNA molecule, and the DNA undercoiling would result from the fact that the double helix spontaneously unwinds under tension. The hRad51 protein adapts to the topology of the dsDNA molecule by switching between the ADP-bound *C-state* and the ATP-bound *S-state*.

According to the above, the NpF works as a mechanical sensor, in which the ATP/ADP conversion is mechanically triggered above/below a critical energy (or torque). The ATP hydrolysis is not used to bias a mechanical transition, as typically happens in molecular motors, but it is used to dissipate the energy and to trigger the irreversibility of a chemical exchange after homology recognition. More specifically, the NpF bears many similarities with the F1-ATPase/synthase, a quasi-reversible molecular motor. In fact, the dsDNA molecule acts as a rotor inside the hRad51 NpF, like the γ-subunit does in the F1-ATPase. In both cases a counterclockwise rotation of this central pivot (the dsDNA molecule or the γ-subunit) triggers a change in the chemical state of the complex from ATP to ADP, and vice-versa for a clockwise rotation. This similar behavior is indeed also associated with striking structural similarities. In particular, the F1-ATPase is a ring constituted of three heterodimers and the NpF is a helix constituted of three homodimers per turn[28]. Interestingly, in the absence of DNA the recombinase also assemble in rings, suggesting that the helical geometry of the NpF is dictated by the presence of the DNA[29,30]. The loops that bind DNA in recombinase proteins are topologically similar to those that bind the coiled-coil γ-subunit in the F1-ATPase[31]. Eventually, the nucleotide-binding sites of the two proteins are structurally homologous[32]. Due to the proximity between the L1 loop and the ATP-binding site in hRad51, the ATP binding can naturally lead to changes of filament architecture resulting in the intercalation-like insertion into the dsDNA[33,34] and in the consequent stretching.

In conclusion, we measured the dynamics and the thermodynamics of a nucleoprotein filament assembled on dsDNA molecules, during its transitions between the stretched and the condensed forms. Our observations suggest that, in the absence of external mechanical constraints, the conformational and the chemical states of hRad51 proteins are strongly correlated (*S* = ATP; *C* = ADP). We also prove that the switch between *S* and *C* states is mechanically triggered by an external torsion induced by the magnetic tweezers. We evaluate that a similar torsional stress is occasioned by the dsDNA molecule, suggesting that hRad51 proteins only stretches the DNA molecule. The DNA undercoiling would result from the spontaneous unwinding of a DNA double helix under tension, and the hRad51 protein adapts to the topology of the dsDNA molecule by switching between the ADP-bound *C-state* and the ATP-bound *S-state*. However, we measure that the simple interaction between the hRad51 helix and the two DNA molecules does not release enough energy to stretch, unwind and denature the DNA fragments. The additional chemical energy released during the pairing of homologous bases is required to stretch the dsDNA molecules, to occasion the concomitant transition of the nucleoprotein filament into the C-*state* and to trigger its disassembly. Unlike conventional molecular motors, the recombinase proteins work as a non-linear mechanical sensor and use the ATP hydrolysis to make the reaction irreversible when the homologous DNA sequence is found, by dismantling the nucleoprotein filament. Eventually, we provide a physical mechanism to explain how the little free energy gain, due to the pairing between homologous sequences, is sufficient to trigger the irreversibility of the HR process.

## Table

| Parameter | Description | Value | Reference |
|---|---|---|---|
| $K_\ell$ | NpF elastic constant (per hRad51) | 800 pN/nm | [17] |
| $\ell_C$ | NpF pitch in the *C-state* (per hRad51) | 1.02 nm | *3 bp of B-DNA*[35] |
| $\Delta\ell$ | NpF pitch increase from *C* to *S-state* (per hRad51) | 0.53 nm | [4] |
| **$K_\vartheta$** | NpF torsional stiffness (per hRad51) | **To be determined** | **Fitted** + [16] |
| $\vartheta_C$ | NpF torsion in the *C-state* (per hRad51) | 1.81 Rad | *3 bp of B-DNA* |
| $\Delta\vartheta$ | NpF torsion increase from *C* to *S-state* (per hRad51) | - 0.78 Rad | [16] |
| **$\Delta\mu_{C\to S}$** | Energy difference between *C* and *S-state* (per hRad51) | **To be determined** | **Fitted** |
| $F$ | External force (applied by the tweezers) | 0-10 pN | *Measured* |
| $L$ | Measured length of the DNA molecule | 0-8 μm | *Measured* |
| $\Gamma$ | External torque (applied by the tweezers) | 0-30 pN·nm | *Measured* |
| $\Theta$ | Measured torsion of the DNA molecule | 5000-10000 Rad | *Measured* |

**Table 1: List of the parameters introduced in equation (1), which describes the two-state elastic model . In bold, the undetermined ones.**

# Methods

## Production and purification of Rad51 protein

Human Rad51 gene was inserted at the NdeI site of the pET15b expression vector (Novagen) and expressed in the Escherichia coli JM109 (DE3) strain that also carried an expression vector for the minor transfer RNAs [Codon(+)RIL®, Novagen]. The protein was purified on Nickel-nitrilotriacetic acid agarose (Invitrogen, France). The hexahistidine tag was then removed from the human Rad51 protein sequence by incubation with 1.5 units of thrombin protease (Amersham Biosciences) per mg of Rad51 during 18 h. The tag-free protein was further purified by chromatography on a MonoQ column (Amersham Biosciences). The Rad51-containing fractions were dialysed against storage buffer (20mM Tris–HCl, pH 8, 0.25mM ethylenediaminetetraacetic acid (EDTA), 20% glycerol, 5mM dithiothreitol (DTT) and 200mM KCl) and kept at -80ºC. Protein concentrations were determined using the Bio Rad protein assay kit with bovine serum albumin (Pierce) as a standard.

## DNA construction

Two different DNA constructs were used. The first DNA construct was composed of a 14435bp central fragment ligated at one end to a multidigoxigenin-labelled DNA fragment of 672bp and at the other end to a multibiotin-labelled fragment of 834±81bp. All fragments were obtained by polymerase chain reaction of the λ-phage DNA; the central fragment was amplified between the positions 22180 and 37096 (Expand Long Template PCR System, Roche) and digested by MluI and EagI (New England Biolabs) at respectively 22220 and 36654; the biotin fragment was amplified between positions 35901 and 37568 with biotin-modified dUTP (Roche) and digested by EagI at 36554; the digoxigenin fragment was amplified between positions 20281 and 20962 with digoxigenin-modified dUTP (Roche) and digested by MluI at 20952.

The second DNA construct was composed of a 10338bp central fragment ligated at one end to a multidigoxigenin-labelled DNA fragment of 907±172bp and at the other end to a multibiotin-labelled fragment of 696±496bp. The central fragment was obtained by transformation of the pREP4 vector in E.Coli cells (Turbo Competent *E.Coli* cells, New England Biolabs) which was linearized through digestion by HindIII and NotI (New England Biolabs) at respectively positions 592 and 603. The digoxigenin and biotin fragments were obtained by PCR with digoxigenin-modified or biotin-modified dUTP (Roche). The digoxigenin fragment was amplified between positions 43063 and 44875 of the λ-phage DNA and digested by HindIII at 44141; the biotin fragment was amplified between positions 4557 and 5947 of the pTYB4 vector and digested by NotI at 5748.

PCR products were purified on spin columns (BD Chroma Spin 1000 or 100) and fragment ligation (T4 DNA ligase, New England Biolabs) was conducted with excess multidigoxigenin and multibiotin fragments to ensure optimal reaction of most of the central fragments. Ligation products were then purified and selected through gel extraction (QiaQUICK Gel extraction kit, QIAGEN). The final products were unnicked DNA molecules of respectively 5.42±0.03µm and 4.06±0.23µm with multiple biotin labels on one end and multiple digoxigenin labels on the other end.

## Microfluidic setup

The biotin-labeled ends of DNA molecules were bound to streptavidin-coated 2.8µm magnetic beads (Dynabeads® M-280 Streptavidin) in a binding buffer (10mM Tris–HCl, pH 7.5, 1mM EDTA, 50mM NaCl) by interaction of the biotin label with the streptavidin. The DNA-bound bead suspension was then introduced at a controlled flow rate into a polydimethylsiloxane (PDMS) microchannel. After 30 min of incubation, most of the unbound beads were washed out of the channel with TE buffer (10mM Tris–HCl, 1mM EDTA, pH 7.5). Flow chamber, a PDMS microchannel 2cm x 2mm x 110µm (total volume: 4.4µl), was placed on a glass coverslip of 24x40mm (Erie Scientific Company, France) treated with Sigmacote® (Sigma-Aldrich) followed by anti-digoxigenin (Roche, France) for subsequent binding of digoxigenin-labelled DNA molecules. Before first use of the channel, Pluronic F-127® (Sigma-Aldrich) was injected into it and incubated overnight at 4ºC to minimize adsorption of hRad51 onto the glass surface and onto the PDMS walls.

## Setup description

Custom hybrid magnetic tweezers were used with a configuration including a centered main ring magnet coupled with auxiliary cylindrical magnets on the sides. The main magnet was composed of the stacking of three ring magnets of 6x2mm with holes of 2mm in diameter (R-06-02-02-G, Supermagnete). The auxiliary magnets were composed of two 4x7mm cylindrical magnets (S-04-07-N, Supermagnete) which were placed on either side of the main ring magnet with opposite polarities. The use of five step by step motors allowed the user to control the XYZ position of the main magnet, the auxiliary magnets' rotation and their height relatively to the main magnet. The main ring magnet applied a vertical magnetic field with a vertical gradient, pulling the beads upwards without hindering their rotation. The auxiliary magnets could be used to apply null to increasingly strong horizontal magnetic fields. The magnetic torques thus applied could then be measured through the analysis of the beads' angular Brownian motion.

## Torque Calibration

The experimental setup used here was an improvement of the one proposed by Lipfert et al.[36]. It aims at easily switching between "classical" and "free-orbiting"[37] magnetic tweezers. The first are useful to impose the supercoiling degree, while the second ones allow a free rotation of the beads around the vertical axis. The setup is constituted by a main hollow cylindrical magnet (white in Figure 1a) coupled with a pair of side cylindrical magnets (black). The cylinder imposes a vertical gradient of the magnetic field that depends on its distance from the bead. This gradient is used to pull the bead up with a defined force. The side magnets can be raised or lowered at will. When lowered, the smaller magnets add a horizontal component to the magnetic field that hinders rotations of the beads. If raised, the contribution of the smaller magnets is negligible and thus equivalent to free-orbiting magnetic tweezers.

The torque applied by the magnetic tweezers is calibrated by monitoring the angular fluctuations of the beads. Through the Boltzmann equation, we experimentally deduce the potential energy of the bead as a function of its angular position (Figure SI2A). For small angular fluctuations, the potential is approximatively harmonic. By fitting the power spectrum of the bead fluctuations to a Lorentzian function (Figure SI2B), we simultaneously evaluate the angular stiffness of the tweezers and the viscous drag of the bead. The calibration is repeated for each bead and for different heights of the magnets, before or after the experiment.

## Aknowledgements

We thank H. Arata for conceiving the first experimental setup, A. Renodon-Cornière and M. Takahashi for providing the hRad51 proteins and T. Strick for his valuable help to produce the DNA fragments. We also warmly thank P. Martin and J. Prost, who carefully read the manuscript and provided critical comments. GC dedicates this article to Valerio Cappello, with whom he wrote the last chapter of the story.

## Funding

Agence nationale de la recherche (ANR-10-BLAN-1013 1 ‑'DynRec'); Region Ile-de-France in the framework of DIM Centre de compétences en nanosciences en Ile-de-France (Project : RecombinaTarget'). Program «Investissements d'Avenir» ANR-10-LABX-31, ANR-10-EQUIPX-34 and ANR-10-IDEX-0001-02 PSL.

## Authors Cotribution

A.D. developed the first version of torque-sensitive magnetic tweezers and performed the initial experiments. S.X.A. redesigned the experimental setup and performed most of the biochemical and biophysical experiments described in the article. D.M. contributed with experiments on the assembly of hRad51 on unconstrained dsDNA molecules. G.C. analyzed the data and developed the theoretical model. J.L.V. and G.C. designed the experiments and wrote the manuscript.